\documentclass[twocolumn,showpacs,amssymb,amsmath,aps,prb]
{revtex4}

\newcommand{\half}{\mbox{$\frac{1}{2}$}}
\usepackage{graphics}
\usepackage{bm}
\usepackage{amsmath}
\usepackage{amssymb}
\begin{document}

\title{Separability and entanglement in finite dimer-type chains
in general transverse fields}
\author{N. Canosa, R. Rossignoli, J.M. Matera}
\affiliation{Departamento de F\'isica-IFLP,
Universidad Nacional de La Plata, C.C. 67, La Plata (1900), Argentina}
\date{Febrary 11 2010}

\begin{abstract}
We determine the conditions under which general dimer-type spin chains with 
$XYZ$ couplings of arbitrary range in a general transverse field will exhibit
an exactly separable parity-breaking eigenstate. We also provide sufficient
conditions which ensure that it will be a ground state. We then examine the
exact side limits at separability of the entanglement between any two spins in
a finite chain, showing that in the vicinity of separability, the system will
loose all signatures of dimerization, with pairwise entanglement approaching
infinite range and becoming independent of separation and interaction range.
The possibility of a non-uniform exactly separable ground state induced by an
alternating field is also shown. As illustration, we examine the behavior of
the pairwise entanglement in a finite $XY$ dimer chain under a uniform as well
as alternating field. Related aspects of the magnetization are also discussed.
\end{abstract}
\maketitle

\section{Introduction}
Quantum entanglement is an essential resource for quantum information science,
allowing radically new forms of information transmission and processing
\cite{Be.93,Ek.91,DV.99,NC.00}. It has also aroused great interest in condensed
matter and many-body physics \cite{AOFV.08}, providing a novel perspective for
the analysis of strongly correlated systems. Fundamental properties of
entanglement in quantum spin chains have been determined, especially in
connection with critical phenomena in the thermodynamic limit
\cite{ON.02,VV.03,AOFV.08}. The study of finite spin chains can also provide
new insights into the most basic aspects of entanglement, and is presently also
stimulated by the unprecedented level of control that can be reached in some
recently developed quantum devices \cite{DDL.03,CAB.08}, able to realize spin
arrays with controllable Heisenberg interactions.

A fundamental related question is the range the entanglement between individual
spins can reach under the action of an applied magnetic field. At the standard
critical field of large anisotropic $XY$ or $XYZ$ chains, it remains finite and
typically small (for instance, restricted to just first and second neighbors in
a $1D$ Ising chain in a transverse field\cite{ON.02}). However, it can diverge
at a different field: Anisotropic chains may also exhibit a {\it factorizing}
field, where an exactly {\it separable} ground state (GS) becomes possible,
i.e., where the mean field GS becomes exact. This remarkable feature was first
discovered in $1D$ chains with first neighbor couplings\cite{K.82,MS.85}, and
recently examined in detail in more general systems in a uniform field
\cite{RR.04,DV.05,AA.06,GI.07,RCM.08,GAI.08,GAI.09,GG.09}. A general method for
determining separability was in particular developed in refs.\
\cite{GAI.08,GAI.09}. In the immediate vicinity of the factorizing field, the
pairwise entanglement in a finite chain can reach {\it full
range}\cite{AA.06,RCM.08}. The transverse factorizing field in finite $XYZ$
chains arises actually at the crossing of opposite $S_z$-parity
levels\cite{RCM.08}, with separable parity breaking eigenstates emerging from
the superposition of the {\it entangled} definite parity states \cite{RCM.09}.

The aim of this work is to examine the previous issues in finite dimer-type
arrays, which have recently received much attention
\cite{KSV.07,HXG.08,GAI.08,GG.09}. We will consider arrays of arbitrary spins
with $XYZ$ couplings of arbitrary range in a transverse field, not necessarily
uniform, and determine the separability conditions together with the
entanglement side limits at separability, which will be shown to be independent
of separation, coupling range and other details such as the strength of the
coupling between dimers. At these points all traces of dimerization will then
be lost. We will also examine factorization under an {\it alternating field},
which can give rise to a separability curve with {\it field dependent}
separable solutions and entanglement limits. Entanglement between spins
unconnected by the interaction can in this way exceed that between linked pairs
in the vicinity of separability. These effects are specially noticeable for
finite chains close to the $XXZ$ limit. Related aspects of the magnetization
and the entanglement between one spin and the rest of the chain are also
discussed.

Section II contains the general theoretical results, including the mean
field+RPA interpretation of the separability conditions, while sec.\ III the
application to general dimer-type chains, including illustrative exact results
for finite chains. The appendix contains the details of the exact calculation
obtained through the Jordan Wigner transformation. Conclusions are finally
drawn in IV.

\section{Formalism}
\subsection{Transverse Factorizing Fields}
We first consider the general Hamiltonian
\begin{equation}
H=\sum_i b^i s^z_i-\half\sum_{i\neq j}(v_x^{ij} s_i^xs_j^x+
v_y^{ij}s_i^ys_j^y+v_z^{ij}s_i^zs_j^z)\,,\label{H1}
\end{equation}
which describes an array of $n$ spins $s_i$ not necessarily equal, interacting
through $XYZ$-type couplings of arbitrary range in a general transverse applied
field $b^i$. It satisfies $[H,P_z]=0$, where $P_z=\exp[i\pi \sum_{i=1}^n
(s_i^z+s_i)]$ denotes the global $S_z$ {\it parity or phase-flip} (here $s_i$
is the spin value at site $i$). Denoting with $|0_i\rangle$ the local state
with its spin fully aligned along the $-z$ direction
($s_i^z|0_i\rangle=-s_i|0_i\rangle$), this Hamiltonian will exhibit a fully
{\it separable} parity breaking eigenstate of the form
\begin{eqnarray}
|\Theta\rangle\equiv|\theta_1\ldots\theta_n\rangle=\otimes_{i=1}^n
\exp[i\theta_i s^y_i]|0_i\rangle\,,\label{Th}
\end{eqnarray}
i.e., a state with its spins fully aligned along local axes forming angles
$\theta_i$ with the $z$ axis, if (and only if) the conditions
\begin{eqnarray}
v_y^{ij}&=&v_x^{ij}\cos\theta_i\cos\theta_j+v_z^{ij}\sin\theta_i\sin\theta_j\,,
\label{a}\\b^i\sin\theta_i&=&\sum_{j\neq i}
s_j(v_x^{ij}\cos\theta_i\sin\theta_j-v_z^{ij}\sin\theta_i\cos\theta_j)
\,, \label{b}
\end{eqnarray}
are satisfied\cite{RCM.09}. They can be obtained replacing $s_i^\mu$ in $H$ by
the rotated operators $e^{-i\theta_i s_i^y}s_i^\mu e^{i\theta_is_i^y}$ and
solving $H_\Theta|0\rangle=E_\theta|0\rangle$, where
$H_\Theta=e^{-i\sum_i\theta_i s_i^y}H e^{i\sum_i\theta_is_i^y}$ and
$|0\rangle=\otimes_i |0_i\rangle$. Eqs.\ (\ref{a})--(\ref{b}) actually hold for
general local rotations $e^{i\bm{\phi}_i\cdot\bm{s}_i}|0_i\rangle$ since the
latter can also be cast in the form (\ref{Th}) through complex angles
$\theta_i$ and a suitable normalization factor\cite{RCM.09,ACGT.72}. Note also
that for a spin $1/2$ array Eq.\ (\ref{Th}) is in fact the most general
separable state. The energy $E_\Theta$ becomes
\begin{eqnarray}
E_\Theta&=&-\sum_{i=1}^n s_i [b^i\cos\theta_i+
\half\sum_{j\neq i}s_j(v_x^{ij}\sin\theta_i\sin\theta_j\nonumber\\
&&+v_z^{ij}\cos\theta_i\cos\theta_j)]\,.\label{en}
\end{eqnarray}

If $|v_y^{ij}|\leq v_x^{ij}\,\, \forall\, i,j$ and $\theta_i\in(0,\pi)$
$\forall$ $i$, $|\Theta\rangle$, as well as its degenerate partner state
\[|-\Theta\rangle=P_z|\Theta\rangle=|-\theta_1,\ldots-\theta_n\rangle\,,\]
will be {\it ground states} of $H$ when Eqs.\ (\ref{a})--(\ref{b}) are
fulfilled\cite{RCM.09}. Of course, they can be GS also in other cases
\cite{K.82,MS.85} by suitably adjusting the relative signs of the $\theta_i$
(see sec.\ \ref{3a}).
Eqs.\ (\ref{b}) are in fact the stationary conditions for the energy (\ref{en})
at fixed $b^i$, $v_\mu^{ij}$, representing the mean field equations which
ensure stability of $|\pm\Theta\rangle$ against one-spin excitations.

Eqs.\ (\ref{a}) warrant that $|\pm\Theta\rangle$ will be exact eigenstates by
canceling the residual matrix elements linking $|\pm\Theta\rangle$ with
two-spin excitations, and have a clear meaning within the {\it random phase
approximation}\cite{RS.80,MRC.08} (RPA): If satisfied, $\forall i,j$ the RPA
vacuum {\it will coincide with the mean field state}. More explicitly, the zero
temperature RPA matrix (whose eigenvalues are the RPA energies) is
\begin{eqnarray}
{\cal H}_{\rm RPA}&=&\left(\begin{array}{cc}
A&B^-\\-B^-&-A\end{array}\right)\,,\;\;\;
A_{ij}=\lambda_i\delta_{ij}+B^+_{ij}\,,\nonumber\\
B^\pm_{ij}&=&-\half\sqrt{s_is_j}(v_x^{ij}\cos\theta_i\cos\theta_j
+v_z^{ij}\sin\theta_i\sin\theta_j\pm v_y^{ij})\,,\nonumber
\end{eqnarray}
where $\pm\half \lambda_i$ are the eigenvalues of the local mean field
Hamiltonian $b^is^z_i-\sum_{j,\mu} v^{ij}_\mu\langle s_j^\mu\rangle_{\Theta}
s_i^\mu$, and $B^\pm_{ij}$ the elements associated with the dispersion ($s_i^+
s_j^-$) and creation ($s_i^+ s_j^+$) of spin excitations respectively. Eq.\
(\ref{a}) is then equivalent to the condition $B^-=0$, implying no RPA
corrections to the mean field vacuum.

From Eq.\ (\ref{a}) it is seen that a {\it uniform} eigenstate with
$\theta_i=\theta$ $\forall i$ becomes feasible if the anisotropy ratio
\begin{equation}
\chi\equiv\frac{v_y^{ij}-v_z^{ij}}{v_x^{ij}-v_z^{ij}}=
\cos^2\theta\,,\label{chi}
\end{equation}
is {\it constant} for all pairs and satisfies $\chi>0$ (if $\chi>1$ (complex
$\theta$), a global rotation of $\pi/2$ around the $z$ axis will lead to
$\chi\rightarrow 1/\chi$ and $\theta$ real). Eq.\ (\ref{b}) leads then to
\begin{equation}
b^i=\sqrt{\chi}\sum_{j\neq i}(v_x^{ij}-v_z^{ij})s_j\,.
 \label{bs}\end{equation}
if $\chi\in[0,1)$ (the opposite sign for all $b^i$ is obviously also feasible)
and to $b^i$ arbitrary if $\chi=1$ ($XXZ$ or $XX$ case\cite{CR.07}, where
$|0\rangle$ is an exact eigenstate $\forall$ $b^i$). Any spin array with
couplings satisfying Eq.\ (\ref{chi}) will then exhibit a uniform separable
degenerate eigenstate $|\Theta\rangle=|\theta,\ldots,\theta\rangle$ if the
fields $b^i$ at sites $i$ are tuned at the values (\ref{bs}). It will be a GS
when $|v_y^{ij}|< v_x^{ij}$ $\forall i,j$.

\subsection{Entanglement at factorizing fields}
In a finite array the exact GS of $H$ will not be in general exactly degenerate
away from the factorizing point, and will have a {\it definite $S_z$-parity}.
The correct side limits at the factorizing field are then provided by the
normalized definite parity states
\begin{eqnarray}
|\Theta^\pm\rangle&=&\frac{|\Theta\rangle\pm|-\Theta\rangle}
 {\sqrt{2(1\pm\langle-\Theta|\Theta\rangle)}}\label{Ps},\end{eqnarray}
where
$\langle-\Theta|\Theta\rangle={\textstyle\prod_{i=1}^n}\cos^{2s_i}\theta_i$ is
the overlap between the degenerate separable eigenstates. The states (\ref{Ps})
satisfy $P_z|\Theta^{\pm}\rangle=\pm |\Theta^{\pm}\rangle$ and are obviously
also exact eigenstates when Eqs.\ (\ref{a})-(\ref{b}) are fulfilled.

These states are entangled, with Schmidt rank\cite{NC.00} 2 for {\it any}
bipartition $(A,\bar{A})$ of the whole system\cite{RCM.09} (here $A$ denotes a
subset of spins and $\bar{A}$ the complementary subset). Moreover, the reduced
state of any subsystem of two or more spins can be effectively considered as a
two-qubit mixed state with respect to any bipartition\cite{RCM.09}. The
entanglement between any two subsystems can then be measured through the {\it
concurrence}, a measure of entanglement originally introduced for two qubit
systems\cite{WW.98} (where it can be exactly computed, see sec.\ \ref{3d}), and
later extended to mixed states of general bipartite systems through the convex
roof extension of the generalized pure state expression\cite{RC.01,RC.03}. The
concurrence between any two spins $i,j$ in the states $|\Theta^\pm\rangle$ can
be shown to be\cite{RCM.09}
\begin{equation}
C^\pm_{ij}=\frac{\sqrt{(1-\cos^{4s_i}\theta_i)(1-\cos^{4s_j}\theta_j)}
\langle-\Theta_{\overline{ij}}|\Theta_{\overline{ij}}\rangle}
 {1\pm\langle-\Theta|\Theta\rangle} \label{Cij}\,,\end{equation}
where $\langle-\Theta_{\overline{ij}}|\Theta_{\overline{ij}}\rangle=
\prod_{k\neq i,j}\cos^{4k}\theta_k$ denotes the complementary overlap. It will
be appreciable for sufficiently small angles $\theta_k$ if
$\theta_i,\theta_j\neq 0$.

On the other hand, the entanglement between one spin and the rest of the chain
in the states (\ref{Ps}) can be measured through the entropy $S_i=-{\rm
Tr}[\rho_i\log\rho_i]$, where $\rho_i= {\rm
Tr}_{\,\bar{i}}\,|\Theta^\pm\rangle\langle\Theta^\pm|$ is the reduced density
matrix of the spin at site $i$, or alternatively, through the pure state
concurrence\cite{RC.01} $C_{i}=\sqrt{2(1-{\rm Tr}\rho_i^2)}$. The latter
provides an upper bound to the sum of all pairwise concurrences $C_{ij}$
stemming from site $i$\cite{CKW.00,ON.06,RCM.09}: $C^2_{i}\geq \sum_{j\neq
i}C_{ij}^2$. It fully determines $S_i$ when $s_i=1/2$ (sec.\ \ref{3d}). Its
expression in the states (\ref{Ps}) reads\cite{RCM.09}
\begin{equation}
C^\pm_{i}=\frac{\sqrt{(1-\cos^{4s_i}\theta_i)
(1-\prod_{k\neq i}\cos^{4s_k}\theta_k)}}{1\pm
\langle -\Theta|\Theta\rangle}
 \,.\label{Ci}
\end{equation}
with $C^\pm_i\approx\sqrt{1-\cos^{4s_i}\theta_i}$ if the overlap is neglected.
The entanglement between $L$ and $n-L$ spins, as well as between any two sets
of spins in the states (\ref{Ps}) can also be exactly calculated\cite{RCM.09}.

When $|\pm\Theta\rangle$ are GS, Eqs.\ (\ref{Cij})--(\ref{Ci}) represent the
actual side limits of the GS concurrences $C_{ij}$ and $C_{i}$ at the
factorizing point, where a transition
$|\Theta^+\rangle\rightarrow|\Theta^-\rangle$ will take place as the field
increases\cite{RCM.08,RCM.09}. The entanglement between two spins will then
reach {\it full range} in its vicinity, provided
$\langle-\Theta_{\overline{ij}}|\Theta_{\overline{ij}}\rangle \neq 0$ and
$\theta_{i}\neq 0$, $\forall$ $i,j$ (Eq.\ (\ref{Cij})).

When $\theta_i=\pi/2$ $\forall i$ (corresponding for $v_z=0$ to the Ising case
$v_y=0$ according to Eq.\ (\ref{chi})), $|\pm\theta_i\rangle$ are orthogonal
and $C_i^\pm=1$ while $C^\pm_{ij}=0$ $\forall$ $i,j$. The previous effect
becomes significant in the opposite limit of small $\theta_i$ (systems close to
the $XXZ$ limit). We also remark that the uniform mixture of both definite
parity states, $\rho_0=\half(|\Theta^+\rangle\langle\Theta^+|
+|\Theta^-\rangle\langle\Theta^-|)$, is also entangled and leads to attenuated
concurrences\cite{RCM.09} $C^0_{ij}=C^-_{ij}\frac{
\langle-\Theta|\Theta\rangle}{1+\langle -\Theta|\Theta\rangle}$,
$C^0_i=C^-_i\frac{\langle-\Theta|\Theta\rangle} {1+\langle
-\Theta|\Theta\rangle}$.

\section{Application to dimer-type chains}
\begin{figure}[t]
\centerline{
\scalebox{.7}{
\includegraphics{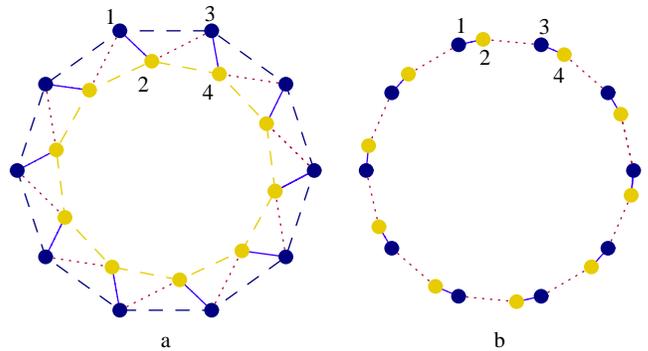}}}
\caption{(Color online) Schematic plot of a) a system described by the
couplings (\ref{vd}), representing two interacting cyclic chains and b) the
dimer chain of Eq.\ (\ref{Hd}), a particular case of a). Numbers
indicate the notation used in figs.\ \ref{f2}--\ref{f3}. }
\label{f1}
\end{figure}

Let us now consider a pair of uniform interacting chains of the same size $m$
and spins $s_\sigma$, not necessarily equal. We can embed this system in a
single non-uniform chain of even size $n=2m$ assigning odd (even) sites to the
first (second) chain, as schematically depicted in Fig.\ \ref{f1} (left), such
that $\sigma=o,e$. We may then consider
\begin{equation}
v_\mu^{ij}=v_\mu^{\sigma_i}(j-i)\,,\label{vd}
\end{equation}
where $\sigma_i=o,e$ indicates the parity of the site, such that
$v_\mu^\sigma(l)$ represents the interchain (internal) couplings for $l$ odd
(even). Accordingly, $v^o(l)=v^e(-l)$ for $l$ odd and
$v^\sigma(l)=v^\sigma(-l)$ for $l$ even (but $v^o(l)\neq v^e(l)$ in general).
In the cyclic case  $v^{\sigma}(-l)=v^{\sigma}(n-l)$ $\forall$ $\sigma,l$.

An important example of this type is that of a  {\it dimer chain} with just
nearest neighbor couplings (Fig.\ \ref{f1}, right), where
$v_\mu^\sigma(l)=v_\mu^\sigma \delta_{l,\pm 1}$:
\begin{equation}
H_d=\sum_{i=1}^{n/2}[b^{2i-1} s^z_{2i-1}+b^{2i} s^z_{2i}
-\sum_{\mu=x,y,z}v^o_\mu s^\mu_{2i-1}s^\mu_{2i} +v^e_\mu
s^\mu_{2i}s^\mu_{2i+1}]\,.
 \label{Hd}\end{equation}
Here $v^e_\mu$ can be considered as the (weak) couplings between dimers and
$v^o_\mu$ the (strong) internal couplings, the system becoming dimerized (i.e.,
an array of independent spin pairs) for $v^e_\mu\rightarrow 0$ (see also sec.\
III C).

A different example of (\ref{vd}), which nonetheless will exhibit factorization
properties similar to those of Eq. (\ref{Hd}) (see below), is a pair of arrays
with no internal couplings interacting through a constant full range coupling:
$v_\mu^{\sigma}(l)=2v_\mu/n$ $\forall$ $l$ odd and $0$ otherwise, such that
\begin{equation}
H_p=b^oS^z_o+b^eS^z_e-{\textstyle\frac{1}{n}}\sum_{\mu=x,y,z}v_\mu S^\mu_o
 S^\mu_e\,,
 \label{Hl}\end{equation}
where $S^\mu_{o,e}=\sum_{l\;{{\rm odd}\atop {\rm even}}}s^\mu_{l}$ are the
total spin components of each array and we have assumed a constant field is
applied to each of them. This system is obviously equivalent to an interacting
pair of spins $S_o=\half n s_o$ and $S_e=\half n s_e$ if restricted to the
maximum spin multiplet. As in the Lipkin model\cite{LMG.65}, the $1/n$ scaling
ensures here a bounded intensive energy $\langle H\rangle/n$ for
$n\rightarrow\infty$ and fixed $v_\mu$.

\subsection{Uniform separable eigenstates \label{3a}}

In the general case (\ref{vd}) with cyclic boundary conditions, a separable
eigenstate with a common angle $\theta_i=\theta$ $\forall i$ is then feasible
if Eq.\ (\ref{chi}) holds for any connected pair, i.e.,
$\chi=\frac{v^\sigma_y(l)-v^\sigma_z(l)}{v^\sigma_x(l)-v^\sigma_z(l)} \in[0,1)$
and constant $\forall\,l$, and there is a uniform field $b^i=b^{\sigma_i}$ in
each subchain given by
\begin{eqnarray}
b^\sigma&=&\sqrt{\chi}\sum_{\sigma'=o,e}v^{\sigma\sigma'}s_{\sigma'}\,,\;\;
\sigma=o,e\,,
 \label{bsx}\end{eqnarray}
where $v^{\sigma\sigma}=\sum_{l\;{\rm even}}v^\sigma_x(l)-v_z^\sigma(l)$,
$v^{oe}=\sum_{l\;{\rm odd}}v^\sigma_x(l)-v_z^\sigma(l)=v^{eo}$, with $b^o=b^e$
if $s_o=s_e$. Such uniform eigenstate is also feasible for a similar {\it open}
chain provided a non-uniform field, as determined by Eq.\ (\ref{bs}), is
applied. For short range couplings this will imply just boundary corrections.
The ensuing states $|\pm\Theta\rangle$ will be GS if
$|v^\sigma_y(l)|<v^\sigma_x(l)$ $\forall$ $\sigma,l$.

The definite parity states (\ref{Ps}) will then lead to a finite concurrence
(\ref{Cij}) for {\it any} spin pair, which will depend  on the parity of the
sites but not on their separation: The odd-odd ($C^\pm_{oo}$), even-even
($C^\pm_{ee}$) and odd-even ($C^\pm_{oe}$) concurrences will be given,
according to Eq.\ (\ref{Cij}), by
\begin{eqnarray}
C^\pm_{\sigma\sigma}&=&
\frac{(1-\chi^{2s_\sigma})\chi^{S-2s_\sigma}}{1\pm \chi^S}\,,
\;\;C^\pm_{oe}=\sqrt{C^\pm_{oo}C^\pm_{ee}}\,,\label{Cd}
\end{eqnarray}
where $S=\half n(s_o+s_e)$ is the total spin. The range of the entanglement
between two spins will then increase as the factorizing fields (\ref{bsx}) are
approached in each subchain, becoming {\it independent} of the coupling range
and separation. If $s_o=s_e$, obviously $C_{oo}=C_{ee}=C_{oe}$.

$C^\pm_{\sigma\sigma}$ will be appreciable for sufficiently small $XY$
anisotropies: If $\chi\approx 1-\delta/S$ then $\chi^S\approx e^{-\delta}$ for
small $\delta/S$. In fact, for $\delta\rightarrow 0$ ($XX$ limit)
$C^+_{\sigma\sigma}\rightarrow 0$  but $C^-_{\sigma\sigma}\rightarrow
2s_\sigma/S$ (i.e., $2/n$ for $s_o=s_e$, which is the maximum attainable value
when all pairs are equally entangled\cite{KBI.00}), as 
$|\Theta^+\rangle\rightarrow |0\rangle$ but $|\Theta^-\rangle$ approaches the 
entangled $W$-type state\cite{DVC.00} $\propto \sum_i s^+_i|0\rangle$. In the 
opposite limit (Ising case $\chi=0$), $|\pm\Theta\rangle$ become orthogonal and  $C_{\sigma\sigma}=0$.

In the dimer chain (\ref{Hd}), the uniform separable eigenstate becomes then
feasible if there is a {\it common} anisotropy
$\chi=\frac{v_y^\sigma-v_z^\sigma}{v_x^\sigma-v_z^\sigma}\in[0,1)$ for
$\sigma=o,e$, and the fields are chosen as (Eq.\ (\ref{bsx}))
\begin{equation}
b^o=\sqrt{\chi}v^{oe}s_{e},\;\;b^e=b^os_o/s_e\,,  \label{bdim}
\end{equation}
where $v^{oe}=v^o_x+v^e_x-v^o_z-v^e_z$. In an {\it open} chain we should just
add the {\it border corrections} $b^1=\half b^o$, $b^{n}=\half b^e$ according
to Eq.\ (\ref{bs}). Thus, in the ferromagnetic-type case  $|v_y^\sigma|\leq
v_x^\sigma$ for $\sigma=o,e$, its GS will become {\it uniform} at the
factorizing fields (\ref{bdim}), {\it regardless of the ratio} $v^e_x/v^o_x$
(as long as it is non-zero), loosing there all signatures of a dimerized
structure and leading to the full range concurrences (\ref{Cd}) as side-limits.

Let us also remark that for the nearest neighbor couplings of (\ref{Hd}), the
antiferromagnetic case $v_x^\sigma<0$ $\forall\sigma$ can be brought back to
the previous case by means of local rotations of angle $\pi$ around the $z$
axis at even sites (implying $s^\mu_i\rightarrow (-1)^{i+1}s^\mu_i$ and hence
$v_\mu^\sigma\rightarrow -v_\mu^\sigma$ $\forall$ $\sigma$ for $\mu=x,y$). A
uniform separable eigenstate
$|\!\!\nearrow\nearrow\ldots\rangle\equiv|\theta\theta\ldots\rangle$ in the
rotated system corresponds then to an {\it alternating} solution
$\theta_i=(-1)^{i+1}\theta$ (Ne\'el-type state
$|\!\!\nearrow\nwarrow\nearrow\nwarrow\ldots\rangle\equiv
|\theta,-\theta,\theta,\ldots\rangle$) in the original system. Note that this
holds for arbitrary spins $s_\sigma$ (equal or distinct). Separability (but not 
entanglement) in the $s_o=s_e=1/2$ dimer chain was discussed in \cite{GG.09}, 
with the correct treatment for general antiferromagnetic couplings discussed in 
detail in \cite{GAI.08,GAI.09}.

For {\it even} $m=n/2$ (to avoid frustration effects\cite{GAI.09}), the {\it
mixed} case $v_x^o>0$, $v_x^e<0$ (or viceversa) can also be recast as a
ferromagnetic case $v_x^\sigma>0$ $\forall \sigma$ by means of local rotations
of $\pi$ around the $z$ axis in even sites of both subchains
($s^\mu_{2i+k}\rightarrow(-1)^{i+1}s^\mu_{2i+k}$ for $k=-1,0$ and $\mu=x,y$,
implying $v_\mu^e\rightarrow -v_\mu^e$). The uniform solution corresponds here
to $\theta_{2i+k}=(-1)^{i+1}\theta$ for $k=-1,0$ in the original mixed system,
i.e.\cite{GAI.09,GG.09}
$|\!\!\nearrow\nearrow\nwarrow\nwarrow\ldots\rangle\equiv
|\theta,\theta,-\theta,-\theta,\theta,\ldots\rangle$. Hence, for even $n/2$ we
may always assume $v_x^\sigma>0$ $\forall \sigma$ in (\ref{Hd}).

In the system (\ref{Hl}), the same uniform separable eigenstate becomes
feasible if $\chi= \frac{v_y-v_z}{v_x-v_z}\in[0,1)$ and the fields are set at
the values (\ref{bdim}), with $v^{oe}=v_x-v_z$ (Eq.\ (\ref{bsx})).  We may
again assume $v_x\geq 0$ since its sign can be changed replacing
$S^\mu_e\rightarrow -S^\mu_e$ for $\mu=x,y$. This system will exhibit just
three different GS pairwise concurrences ($C_{oo}$, $C_{ee}$, $C_{oe}$) for any
$b^o$, $b^e$, which will approach the same limits (\ref{Cd}) at the fields
(\ref{bdim}).

\subsection{Alternating separable eigenstates}
In the case of two interacting subchains with no internal couplings, like Eqs.\
(\ref{Hd}) or (\ref{Hl}), we may also consider the possibility of {\it
different} and {\it controllable} uniform angles $\theta_o$, $\theta_e$ (with
$|\theta_o|\neq|\theta_e|$) in each subchain, i.e.,
\[|\Theta\rangle=|\theta_o\theta_e\theta_o\theta_e\ldots\rangle\,,\]
through an alternating field $b^o\neq b^e$.  For simplicity  we will consider
$XY$ couplings ($v_z^{ij}=0$). According to Eqs.\ (\ref{a})--(\ref{b}), such a
solution is feasible if for $\sigma=o,e$ and $l$ odd,
\begin{eqnarray}
\chi&=&\frac{v_y^\sigma(l)}{v_x^{\sigma}(l)}=\cos\theta_o\cos\theta_e,\;\;
b^\sigma=v^{oe}\frac{\sin\theta_{\sigma}}{\tan\theta_{\bar{\sigma}}}
s_{\bar{\sigma}},
\end{eqnarray}
where $v^{oe}=\sum_{l\;{\rm odd}}v_x^\sigma(l)$ and $\bar{\sigma}\neq\sigma$
(i.e.\ $v^{oe}=v_x^o+v_x^e$ in (\ref{Hd}) and $v^{oe}=v_x$ in (\ref{Hl})). This
implies
\begin{eqnarray}
b^o b^e&=&\chi (v^{oe})^2 s_o s_e\,,\label{b12}\\
\cos^2\theta_\sigma&=&\frac{\chi^2+
\tilde{b}^2_\sigma}{1+\tilde{b}^2_\sigma}\,,
\;\;\tilde{b}_\sigma\equiv\frac{b^\sigma}{v^{oe}s_{\bar{\sigma}}}\,.
\label{csg}
 \end{eqnarray}
Hence, for fields $b^o,b^e$ satisfying Eq.\ (\ref{b12}) we obtain a separable
eigenstate with alternating angles $\theta_o$, $\theta_e$ determined by Eq.\
(\ref{csg}). Since one of the fields is now {\it free}, we may adjust in such
system the individual angles and thus the internal ($C^\pm_{oo}$,
$C^{\pm}_{ee}$) and interchain ($C^\pm_{oe}=C^{\pm}_{eo}$) pairwise
concurrences at separability, given now by
\begin{eqnarray}
C^\pm_{\sigma\sigma}&=&
\frac{(1-\chi_\sigma^{2s_\sigma})\chi_\sigma^{S_\sigma-2s_\sigma}
\chi_{\bar{\sigma}}^{S_{\bar{\sigma}}}}{1\pm \chi_{\sigma}^{S_\sigma}
\chi_{\bar{\sigma}}^{S_{\bar{\sigma}}}}\,,\;\;
C^\pm_{oe}=\sqrt{C^{\pm}_{oo}C^{\pm}_{ee}}\,,\label{Cijd}
\end{eqnarray}
where $\chi_\sigma\equiv\cos^2\theta_\sigma$ and $S_\sigma=n s_\sigma/2$. If
$|b^o|>|b^e|$ and $s_o=s_e$, $C^\pm_{oo}<C^\pm_{oe}<C^\pm_{ee}$, despite the
absence of even-even direct coupling ($v^{ee}=v^{oo}=0$). For $\tilde{b}^o=
\tilde{b}^e$ we recover Eqs.\ (\ref{Cd})--\ref{bdim}). The values of 
$C^\pm_{\sigma\sigma'}$ depend now on the ratio $\eta=\tilde{b}^o/\tilde{b}^e$ 
($\tilde{b}_o=\sqrt{\eta\chi}$, $\tilde{b}_e=\sqrt{\chi/\eta}$ when  Eq.\
(\ref{b12}) holds). For $\eta\gg 1$, $\cos\theta_o\rightarrow 1$ but
$\cos\theta_e\rightarrow \chi$, implying that in this limit $C^\pm_{ee}$
remains {\it finite} at the factorizing field, while $C^\pm_{oo}$ and
$C^\pm_{oe}$ {\it vanish}. Note also that $\theta_\sigma$ is a decreasing
function of $\tilde{b}_\sigma$.

In the ferromagnetic case $v_x^\sigma>0$, $\theta_o$ and $\theta_e$ have both
the same sign. For antiferromagnetic couplings $v_x^\sigma<0$ $\forall$
$\sigma$ in the dimer chain  (\ref{Hd}), we would have instead $\theta_o>0$ and
$\theta_e<0$ (or vice-versa), whereas in the mixed case $v_x^ev_x^o<0$,
$|\Theta\rangle=|\theta_o\theta_e,-\theta_o,
-\theta_e,\theta_o\theta_e,\ldots\rangle$, as previously discussed. Border
corrections $b^1=\half b^o$, $b^{n}=\half b^e$ would also apply in an open
dimer chain.

The concurrence between one-spin and the rest of the chain $C_{\sigma_i}\equiv
C_{i}$, will be given at separability by (Eq.\ (\ref{Ci}))
\begin{equation}
C^\pm_{\sigma}={\textstyle\frac{\sqrt{(1-\chi_\sigma^{2s_\sigma})
(1-\chi_\sigma^{2(S_\sigma-s_\sigma)}
\chi^{2S_{\bar{\sigma}}}_{\bar{\sigma}})}} {1\pm\chi_{\sigma}^{S_\sigma}
\chi_{\bar{\sigma}}^{S_{\bar{\sigma}}}}}\,.\label{Cig}
\end{equation}

\subsection{Spin $1/2$ pair}
We may explicitly verify the previous expressions (valid for general $n$) in
the ``two qubit'' case ($s_o=s_e=1/2$, $n=2$), which also represents the
$v^e_\mu\rightarrow 0$ limit in the spin $1/2$ dimer chain (\ref{Hd}). Setting
$v_\pm=\frac{1}{4}(v_x\pm v_y)\geq 0$ and $b_{\pm}=\half(b^o\pm
b^e)$, with $b^o=b^1$, $b^e=b^2$, the eigenstates and energy levels of
Hamiltonian (\ref{H1}) become in this case
\begin{eqnarray}
|\Psi^-_{\pm}\rangle&=&\alpha^-_\mp|\!\!\uparrow\downarrow\rangle\pm
\alpha^-_\pm|\!\!\downarrow\uparrow\rangle\,,\;
E^-_\pm={\textstyle\frac{1}{4}}v_z\mp\sqrt{b_-^2+v_+^2},\label{E1}\\
|\Psi^+_{\pm}\rangle&=&\alpha^+_{\mp}|\!\!\uparrow\uparrow\rangle\pm
\alpha^+_{\pm}|\!\!\downarrow\downarrow\rangle\,,\;
E^+_{\pm}=-{\textstyle\frac{1}{4}}v_z\mp\sqrt{b_+^2+v_-^2},
 \label{E2}\end{eqnarray}
where $(\alpha^\nu_{\pm})^2=\half(1\pm\frac{b_{\nu}}
{\sqrt{b_{\nu}^2+v_{-\nu}^2}})$ and the superscript $\nu=\pm$ indicates the
$S_z$-parity. The GS corresponds to $|\Psi^-_+\rangle$ or $|\Psi^+_+\rangle$,
with $E^-_+$ and $E^+_+$ {\it crossing precisely when the factorizing
conditions (\ref{a})--(\ref{b}) hold}. At this point $|\Psi^\pm_+\rangle$
become the states (\ref{Th}). In particular, for an homogeneous field ($b_-=0$,
$b_+=b$), $E_+^-=E_+^+$ when $b=\half\sqrt{\chi}(v_x-v_z)$ (Eq.\ (\ref{bs})),
whereas for $v_z=0$ they cross when $b^o b^e=\frac{1}{4}\chi v_x^2$ (Eq.\
(\ref{b12})). It is then explicitly verified that the states (\ref{Th}) are the
true side limits at the crossing point, with separability arising just from the
crossing of these two states. Factorization corresponds then to the quantum
critical point of the spin $1/2$ pair. It should be also noticed that
$|\Psi^\pm_+\rangle$ can here be always written as projected states (\ref{Th})
using suitable angles
($\tan^2\half\theta_1=\frac{\alpha^+_-\alpha^-_-}{\alpha^+_+\alpha^-_+}$,
$\tan^2\half\theta_2=\frac{\alpha^+_-\alpha^-_+}{\alpha^+_+\alpha^-_-}$). The
concurrence $C_{12}=\sqrt{2(1-{\rm Tr}\rho_1^2)}$ in the states
$|\Psi^\pm_\nu\rangle$ reads
\begin{equation}
C^\pm_{12}=2|\alpha^\pm_+\alpha^\pm_-|=|v_{\mp}|/\sqrt{b_{\pm}^2+v_{\mp}^2}\,,
 \label{cdim}\end{equation}
and coincides with both general results (\ref{Cij})-(\ref{Ci}) for the present
case ($C_{12}^\pm=|\sin\theta_1\sin\theta_2|/(1\pm\cos\theta_1\cos\theta_2)$).

In the spin $1/2$ dimer chain (\ref{Hd}), Eq.\ (\ref{cdim}) represents the
limit of the concurrence $C_{2i-1,2i}$ for $v_\mu^e\rightarrow 0$. This implies
that its GS will become fully dimerized (i.e., an array of maximally entangled
pairs) at zero field, since in this case $C_{12}^{\pm}=1$ and all eigenstates
$|\Psi^\pm_\nu\rangle$ are Bell states. However, at finite fields,
$C^{\pm}_{12}=1$ only if $b_\pm=0$ ($b^e=\mp b^o$), in which case just half of
the eigenstates remain maximally entangled. For $b^{o,e}>0$, maximum
entanglement ($C_{12}=1$) for $v_\mu^e\rightarrow 0$ will then arise just for
an homogeneous field $b_+=b$ {\it lower than the factorizing field}, i.e., when
the pair GS is antiparallel ($|\Psi^-_+\rangle$).

Let us finally notice that for $\nu=\pm$,
\begin{eqnarray}
\langle s^z_i\rangle_{\nu}&\equiv&\langle \Psi^\nu_+|s^z_i|\Psi^\nu_+\rangle=
-\half(\nu)^{i+1}b_\nu/\sqrt{b^2_\nu+v_{-\nu}^2}\,,
 \label{si}\end{eqnarray}
implying $\langle s^z_1\rangle_+=\langle s^z_2\rangle_+$ but $\langle
s^z_1\rangle_-=-\langle s^z_2\rangle_-$, i.e. {\it opposite} magnetizations for
negative $S_z$-parity (see below).

\subsection{Results\label{3d}}
Figs.\ \ref{f2}--\ref{f3} depict illustrative results for the GS pairwise
concurrence $C_{ij}$ in a finite spin $1/2$ dimer chain described by Eq.\
(\ref{Hd}) with cyclic $XY$ couplings ($v_z^\sigma=0$). We
have set $v_\mu^e=\alpha v_\mu^o$, with $v^\sigma_y=\chi v^\sigma_x$ for
$\sigma=o,e$. Full exact results for finite $n$ can in this case be obtained
through the Jordan-Wigner transformation (see Appendix). In this system
$C_{ij}=C_{1,j-i+1}$ ($C_{2,j-i+2}$) for $i$ odd (even).

The reduced density matrix for a pair of spins $i,j$ will commute with the pair
parity $e^{i\pi(s^z_i+s^z_j-1)}$, being then of the form ($\langle
\ldots\rangle$ denotes here the GS average)
\begin{equation}
\rho_{ij}=\frac{1}{4}+\langle s^z_i\rangle s^z_i+\langle s^z_j\rangle s^z_j
 +4\!\!\sum_{\mu=x,y,z}\langle s^\mu_is^\mu_j\rangle s^\mu_is^\mu_j
 \label{rho}\end{equation}
The GS pairwise
concurrence $C_{ij}$ can then be evaluated as \cite{WW.98} $2\lambda_{\rm max}-
{\rm Tr}\,R$, with $\lambda_{\rm max}$ the greatest eigenvalue of the matrix
$R=4\sqrt{s^y_i s^y_j\rho_{ij} s^y_i s^y_j\rho_{ij}}$, and reads
\begin{eqnarray}
C_{ij}&=&{\rm Max}[C^+_{ij},C_{ij}^-,0]\,,\\
C_{ij}^{\pm}&=&2[{\textstyle |\langle s^x_is^x_j\mp s^y_is^y_j\rangle|
-\sqrt{(\frac{1}{4}\mp\langle s^z_is^z_j\rangle)^2-
\frac{1}{4}\langle s^z_i\mp s^z_j\rangle^2}}]\,,\nonumber
\end{eqnarray}
being parallel\cite{AA.06} (i.e., as in the states
$|\!\!\uparrow\uparrow\rangle+|\!\!\downarrow\downarrow\rangle$)  if
$C_{ij}^+>0$ and antiparallel
($|\!\!\uparrow\downarrow\rangle-|\!\!\downarrow\uparrow\rangle$) if
$C^-_{ij}>0$ (just one of $C^\pm_{ij}$ can be positive). The entanglement of
formation of the pair can then be obtained as\cite{WW.98}
$S_{ij}=-\sum_{\nu=\pm}p_\nu\log_2 p_\nu$, where
$p_\pm=\half(1\pm\sqrt{1-C_{ij}^2})$. $C_{ij}=S_{ij}=0$ ($1$) for a separable
(maximally entangled) pair.

The case of a uniform field $b^o=b^e=b$ is depicted in Fig.\ \ref{f2}. Here
$C_{1j}=C_{2,j+1}$ for $j$ odd. At $b=0$ and for $\alpha=1$ (uniform chain),
there is entanglement just between first neighbors ($C_{12}=C_{23}>0$). For a
small anisotropy $\chi=0.9$, as soon as the ratio $\alpha$ decreases below 1
the concurrence between weakly coupled pairs ($C_{23}$) rapidly decreases (top
panel), vanishing here already for $\alpha\alt 0.74$, whereas $C_{12}$ rapidly
increases, practically reaching saturation for $\alpha=0.25$ (center panel).
Hence, at zero field approximate dimerization is achieved already for low
$\alpha$, the system becoming essentially an array of maximally entangled pairs
in the antiparallel states $|\Psi_+^-\rangle$.

\begin{figure}[t]
\centerline{
\scalebox{.9}{\includegraphics{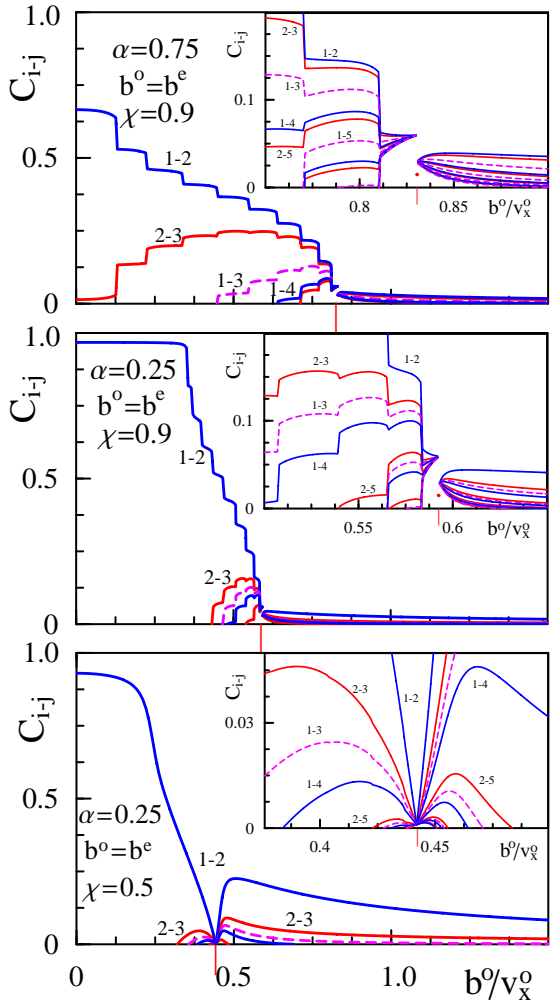}}}

\caption{(Color online) Concurrences between spins $i,j$ vs.\ magnetic field in
a spin $1/2$ $XY$ dimer chain (Eq.\ (\ref{Hd})) for two values of $\alpha\equiv
v^e_\mu/v^o_\mu$ and of the anisotropy $\chi\equiv v^\sigma_y/v^\sigma_x$. The
field is here uniform, with $n=20$ spins. All $C_{ij}$ approach the same side
limits (\ref{Cd}) (which are independent of $\alpha$) at the factorizing field
(\ref{bss}) (red bar), as seen in the insets (blow up of main plot), changing
from antiparallel to parallel and exhibiting there the same finite step. The
red dot at $b=b_s$ indicates the concurrence $C_{ij}^0$ in the mixture of both
definite parity ground states.} \label{f2}
\end{figure}

The previous picture remains valid for weak finite fields. As seen in the top
and central panels, increasing the uniform field destroys dimerization in a
stepwise manner, the GS remaining almost unchanged until the first step,
occurring at $b^*\approx\half\sqrt{\chi}v_x^o(1-\alpha)$ for $\alpha$ not close
to 1. These steps, clearly visible in small chains with low anisotropy, reflect
the $n/2$ GS $S_z$-parity transitions (crossings between the lowest levels of
opposite parity\cite{RCM.08,GG.09}, which are close but not degenerate) taking 
place as the field increases when $\chi\in(0,1]$, as in the homogeneous $XY$ 
chain. At the same time, the concurrence range increases as the last step is 
approached. The latter occurs precisely at the uniform factorizing field 
(Eq.\ (\ref{bs})) 
\begin{equation} b_s=\half\sqrt{\chi}v_x^o(1+\alpha)\,,\label{bss}
\end{equation}
where the dimer structure is completely lost and entanglement reaches full
range: {\it All} pairs become equally entangled irrespective of separation or
location, with $C_{ij}$ reaching the side limits (\ref{Cd}) $\forall$ $i\neq j$
($\lim_{b\rightarrow b_s^{\pm}}C_{ij}=C^\pm_{ij}$), which are {\it independent}
of $\alpha$ and hence {\it the same} in top and central panels. At this field
all $C_{ij}$ exhibit the same finite discontinuity, changing from antiparallel
($b<b_s$) to parallel ($b>b_s$). For $\alpha\rightarrow 0$, $b^*$ and $b_s$
(first and last steps) merge at the two-qubit factorizing field
$\half\sqrt{\chi}v_x^o$.

For stronger fields $b>b_s$ we obtain a weak parallel concurrence, which for
first and second neighbors persists for arbitrarily strong fields and can be
described perturbatively. First (second) neighbors concurrences are first
(second) order in $v_x^\sigma/b$ and given, up to $O(v_x^\sigma/b)^2$, by
\begin{eqnarray}
C_{12}&\approx &|\frac{v_-^o}{b}|-\half(\alpha\frac{v_-^o}{b})^2\,,\;\;
C_{23}\approx |\frac{\alpha v_-^o}{b}|-\half(\frac{v_-^o}{b})^2\,,\nonumber\\
C_{i,i+2}&\approx &|\frac{\alpha v_-^{o}v_+^o}{bb^{\sigma_i}}|
-\half(\frac{v_-^o}{b})^2(1+\alpha^2)\,,
 \nonumber\end{eqnarray}
where $v^o_{\pm}=\frac{1}{4}(v_x^o\pm v_y^o)=\frac{1}{4}v_x^o(1\pm\chi)$ and
$b=\half(b^o+b^e)$. Note that a threshold value of $\alpha$ is required for a
positive second neighbor concurrence for strong fields.

For higher anisotropies (lower $\chi$), the behavior becomes similar to that of
larger systems. The GS parity transitions become less noticeable, as seen for
$\chi=0.5$ in the bottom panel, and the pairwise concurrence side-limits at the
factorizing field are smaller. Nonetheless, the increase of the concurrence
range in its vicinity remains clearly appreciable. Let us remark that for small
separations $|i-j|$, the results for $C^\pm_{ij}$ for $n=20$ at $\chi=0.5$ are
already very close to those for $n\rightarrow\infty$. We should also mention
that as $\chi$ decreases, lower ratios $\alpha$ are required to achieve
approximate dimerization at low fields (at $\chi=0.5$ and $b=0$, $C_{23}$
vanishes only for $\alpha\alt0.58$).

\begin{figure}[t]
\centerline{
\scalebox{.9}{
\includegraphics{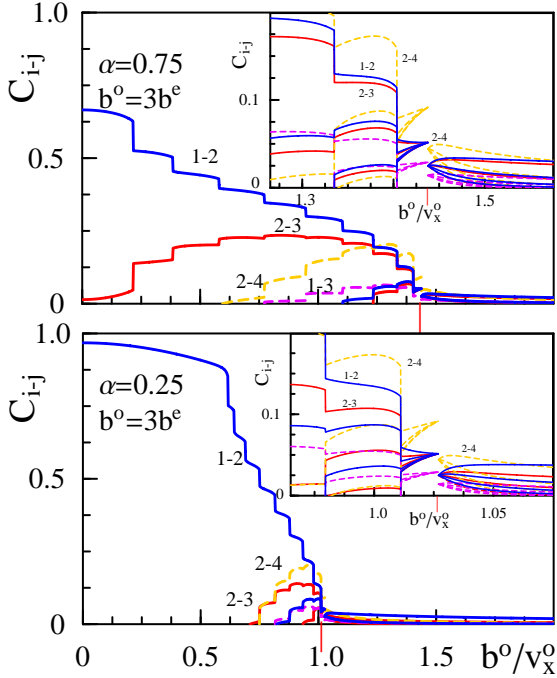}
}
}
\caption{(Color online) Same details as the top and center panel of fig.\
\ref{f2} for the case of different fields at even and odd sites, with a fixed
ratio $b^o/b^e=3$. Now odd-even, odd-odd and even-even concurrences approach
different common side limits at the factorizing field (\ref{bsso}), with
$C_{24}$ becoming the greatest concurrence in its vicinity.} \label{f3}
\end{figure}

Fig.\ 3 depicts the typical behavior for small anisotropies when different
fields are applied at even at odd sites, with a fixed ratio $\eta=b^o/b^e=3$.
The factorizing value for $b^o$ (Eq.\ (\ref{b12})) is here
\begin{equation}
b_{s}^o=\sqrt{\eta}b_s=\half\sqrt{\eta\chi}v_x^o(1+\alpha)\,.\label{bsso}
\end{equation}
At $b_s^o$ there are now three different limits for the concurrences at each
side, $C^{\pm}_{oo}$, $C^\pm_{ee}$ and $C^{\pm}_{oe}$, which represent the
common side limits of $C_{1,2j+1}$, $C_{2,2j+2}$ and $C_{1,2j},C_{2,2j+1}$
$\forall j$ and are given by Eq.\ (\ref{Cijd}). They satisfy here
 \[C^{\pm}_{ee}/C^{\pm}_{oo}=(\chi+\eta)/(\chi+\eta^{-1})>1\]
for $\eta>1$, implying  $C^\pm_{ee}>C^\pm_{oe}$. In particular, $C_{24}$ (which
approaches $C^\pm_{ee}$ for $b^o\rightarrow (b^o_s)^\pm$) clearly exceeds in
the vicinity of $b^o_s$ both first neighbor concurrences $C_{12}$ and $C_{23}$,
despite the absence of second neighbor couplings.

It is also seen that $C_{12}$ is no longer nearly constant up to the first
parity transition, which occurs now at $b^{o*}\approx
\half\sqrt{\eta\chi}v_x^o(1-\alpha)$. This effect can be easily understood with
the two qubit concurrence (\ref{cdim}): At low $\alpha$, $C_{12}$ is
essentially described in the first region by the two qubit expression
(\ref{cdim}) for $C^-_{12}$, which for $b_-=\half b^o(1-1/\eta)\neq 0$, is no
longer constant and decreases as $b^o$ increases.

\begin{figure}[t]
 \centerline{
\scalebox{.75}{\includegraphics{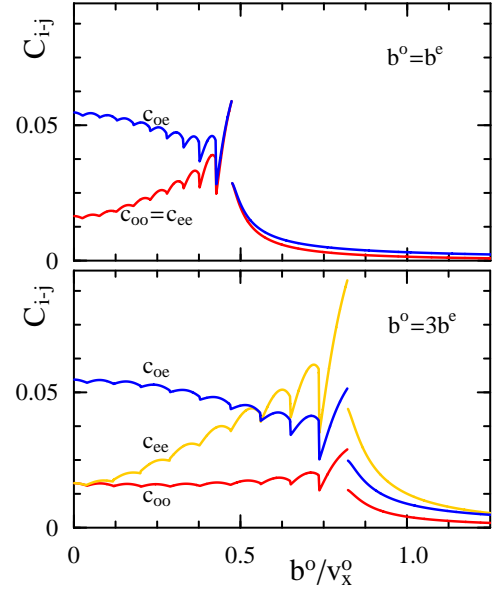}}}
\caption{(Color online) Concurrences between spins $i,j$ vs.\ magnetic field in
a system with constant full range couplings between even and odd sites,
described by Hamiltonian (\ref{Hl}). We have set again $\chi=0.9$, $n=20$ and a
uniform (alternating) field in the top (bottom) panels. Odd-even, odd-odd and
even-even concurrences approach at the factorizing field exactly the same side
limits as those of figs.\ \ref{f2}--\ref{f3} respectively, which are here of
the same order as the values outside this field.} \label{f4}
\end{figure}

In order to highlight the universality of the limits at the factorizing field,
we depict in fig.\ \ref{f4} the pairwise concurrences in the system (\ref{Hl})
for $s_o=s_e=1/2$, where the exact solution can be obtained through direct
diagonalization in the $S_o=S_e=n/2$ representation. In the $XY$ case,
$v^{oe}=v_x$ and the odd factorizing field at fixed ratio $b^o/b^e=\eta$ is
$b^o_s=\half v_x\sqrt{\eta\chi}$. There are now just three different
concurrences at all fields, $C_{oo}=C_{1,2j+1}$, $C_{ee}=C_{2,2j+2}$ and
$C_{oe}=C_{1,2j}=C_{eo}$ ($j$-independent), which approach the same limits of
figs.\ \ref{f2}-\ref{f3} (Eqs.\ (\ref{Cd}), (\ref{Cijd})) at the factorizing
field, since the latter depend solely on $\chi$ and the field ratio $\eta$.
They are here comparable to the values away from the factorizing field, since
the monogamy bound on $\sum_{j\neq i}C_{ij}^2$ entails
$C_{\sigma\sigma'}=O(1/n)$ in this symmetric system\cite{KBI.00}. In the case
considered $C_{oo}$ and $C_{ee}$ are in fact {\it maximum} at the factorizing
field. Note again that for $b^o/b^e>1$, $C_{ee}>C_{oe}$ in the vicinity of
$b^o_s$, despite the absence of even-even couplings.

\begin{figure}[t]
\centerline{
\hspace*{-.5cm}
\scalebox{.69}{\includegraphics{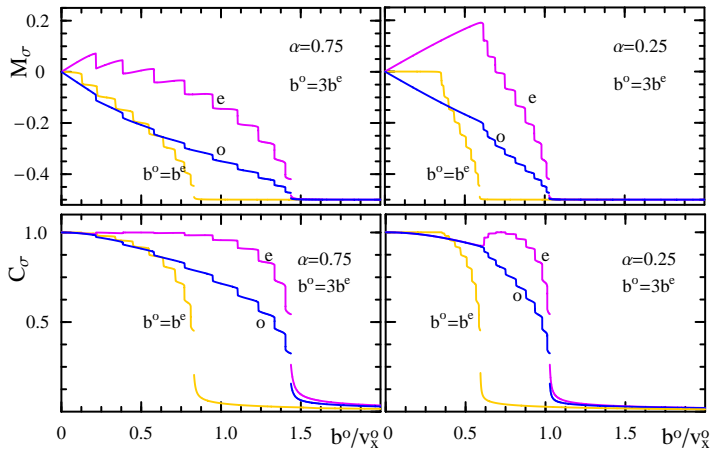}}}
\caption{(Color online) Magnetization at even and odd sites
$M_\sigma\equiv\langle s^z_\sigma\rangle$ (top panels) and the concurrence
$C_\sigma\equiv C_{i}$ (Eq.\ (\ref{Ci2})) between the site and the rest of the
chain (bottom panels), in the dimer chain of figs.\ \ref{f2}--\ref{f3} with
$\chi=0.9$. Results for an alternating field with fixed ratio $b^o/b^e=3$ and
for a uniform field are depicted. The discontinuities at the factorizing field
are explicitly shown. The ``dimer phase'' (fields below the first transition)
presents opposite magnetizations and leads to a non-monotonous behavior of
$C_e$ after this transition. } \label{f5}
\end{figure}

Finally, we depict in fig.\ \ref{f5} the site magnetizations
$M_{\sigma_i}\equiv \langle s^z_i\rangle$ together with the concurrence
$C_{\sigma_i}\equiv C_{i}$ between one spin and the rest of the chain (Eq.\
(\ref{Ci})). For a spin $1/2$ chain with $[H,P_z]=0$, both quantities are
strictly related, since the reduced density matrix for one spin in a state with
definite parity is diagonal in the $s_z$ basis ($\rho_i=\half+2\langle
s_i^z\rangle s_i^z$ as $\langle s_i^\mu\rangle=0$ for $\mu=x,y$) and hence
\begin{equation}
C_{i}=\sqrt{2(1-{\rm Tr}\rho_i^2)}=\sqrt{1-4\langle s^z_i\rangle^2}\,.
\label{Ci2}
\end{equation}
Thus, $C_{i}=1$ when $\langle s^z_i\rangle=0$ (zero field). At the factorizing
field it approaches the side limits (\ref{Ci}). The ensuing entanglement
entropy can be evaluated as $S_i=- \sum_{\nu=\pm} p_\nu\log_2 p_\nu$, with
$p_\pm=\half(1\pm\sqrt{1-C_{i}^2})$.

While for a uniform field the even and odd site magnetizations coincide and
decrease stepwise as the field increases, approaching $-1/2$ for strong fields,
for non-uniform fields they first acquire {\it opposite} signs ($M_o=-M_e$) in
the ``dimer phase'', i.e., before the first parity transition. Here the
magnetization is essentially described by the two-qubit result (\ref{si}),
which yields $M_e=-M_o>0$ in the state $|\Psi_+^-\rangle$ if $b^o>b^e$.
Accordingly, $M_e$ first {\it increases} as $b^o$ (and hence $b^e$ and $b_-$ in
(\ref{si})) increases, in close agreement with Eq.\ (\ref{si}). After the first
transition, $M_e$ starts to decrease, crossing $0$ and approaching $-1/2$
(together with $M_{o}$) for strong fields, even though it may still increase
between transitions.

This entails a non-monotonous behavior of $C_e$ for increasing fields,
particularly appreciable for low $\alpha$, where $C_e$ saturates  again
($C_e=1$) at a {\it finite} field, i.e., when $M_e$ vanishes. At the
factorizing field $C_\sigma$ approaches the limits (\ref{Cig}), which are
independent of $\alpha$, with the magnetization step there given
by\cite{RCM.09} $\Delta M_i\equiv\langle s_i^z\rangle_--\langle
s_i^z\rangle_+=\frac{\sin^2\theta_i
\langle-\Theta_{\bar{i}}|\Theta_{\bar{i}}\rangle}
{1-\langle\!-\Theta|\Theta\rangle^2} $. For strong fields the behavior of
$M_\sigma$ and $C_\sigma$ can again be described perturbatively: We obtain
$M_o\approx M_e\approx -\half[1-(\frac{v_-^o(1+\alpha)}{b})^2]$, with
$C_o\approx C_e\approx |\frac{v_-^o(1+\alpha)}{b}|$.

\section{Conclusions}
We have first determined the factorization conditions for general dimer-type
arrays with XYZ couplings in general transverse fields. We have also examined
the entanglement properties of the associated definite parity states, which
constitute the actual GS side-limits at separability in a finite system,
showing that weak but non-zero full range pairwise entanglement can be reached
in the vicinity of factorizing fields. The possibility of an alternating and
field dependent separable GS through a non-uniform field along a separability
curve (Eq.\ (\ref{b12})) has also been shown, for general spin. Border
corrections to the field allow exact separability also in open chains.

We have then examined the magnetic behavior of a finite spin $1/2$ $XY$ dimer
chain. The factorizing field corresponds to the last parity transition
exhibited by the exact GS for increasing field. Dimerization breakdown takes
then place in steps, with all signatures of dimer structure being completely
lost at the factorizing point: For a uniform field, the concurrence between any
two spins approaches there (at each side) a {\it constant value}, independent
of separation and coupling ratio $v^e/v^o$. The same behavior occurs in an
alternating field, except that in this case there are three different
concurrence side limits at separability, which depend on the odd-even field
ratio. The entanglement between spins unconnected by the coupling may here
exceed that between connected pairs.

The previous properties are not a particular feature of the system considered.
For full range coupling, the same behavior is obtained at separability, as the
eigenstates become there independent of the coupling range. The behavior of the
concurrence between one spin and the rest of the chain has also been examined.
An alternating field can induce {\it opposite} magnetizations at even and odd
sites before the first transition, leading to a non-monotonous behavior of this
concurrence for increasing fields, with two saturation points. The present
results shed light on the complex behavior of entanglement in these systems,
and its relation with factorization. The exposed features can make such finite
critical systems of special interest for diverse applications.

The authors acknowledge support from CIC (RR) and CONICET (NC, JMM) of
Argentina.

\appendix*
\section{Exact solution of the dimer chain with alternating field}
By means of the Jordan-Wigner transformation\cite{LSM.61}, and for a {\it
fixed} value $p=\pm$ of the global $S_z$-parity $P_z$, we may exactly rewrite
the dimer XY Hamiltonian (\ref{Hd}) as a quadratic form in standard fermion
creation and annihilation operators $c^\dagger_j$, $c_j$. For an alternating
field $b^i=b^{\sigma_j}$, with $\sigma_j=o,e$ the site parity, we obtain
\begin{eqnarray}
H_d^{p}&=&\!\!
\sum_{j=1}^n[b^{\sigma_j}
(c^\dagger_jc_j-\half)-\eta^{p}_j(v_{+}^{\sigma_j}c^\dagger_j c_{j+1}
+v_-^{\sigma_j}c^\dagger_j c^\dagger_{j+1}+h.c.)\label{hf1}
\end{eqnarray}
where $n+1\equiv 1$, $v^\sigma_{\pm}=\frac{1}{4}(v_x^\sigma\pm v_y^\sigma)$ and
$\eta^-_j=1$, $\eta^+_j=1-2\delta_{jn}$. By means of separate discrete parity
dependent Fourier transforms for even and odd sites,
\[\left(\begin{array}{c}c^{\dagger}_{2j-1}\\c^\dagger_{2j}\end{array}\right)=
\frac{1}{\sqrt{n}}\sum_{k\in k_{\pm}}
e^{-i\omega k
j}\left(\begin{array}{c}{c'}^\dagger_{k o}\\ {c'}^\dagger_{ke}
\end{array}\right),\;\;
 \omega=\frac{4\pi}{n}\]
where $k_+=\{\half,\ldots,\frac{n}{2}-\half\}$,
$k_-=\{0,\ldots,\frac{n}{2}-1\}$, we may rewrite (\ref{hf1}) as
\begin{eqnarray}H_d^p&=&\sum_{k\in k_{p}}\{\sum_{\sigma}
b^\sigma {c'}^\dagger_{k\sigma}{c'}_{k\sigma}-[v_+^k{c'}^\dagger_{ko}{c'}_{ke}
+v_-^k{c'}^\dagger_{ko}{c'}^{\dagger}_{-ke}+h.c.]\}\nonumber\\
&=&\sum_{k\in k_{p}}\sum_{\nu=\pm}
\lambda_k^\nu (a^\dagger_{k\nu}a_{k\nu}-\half)\,,
 \label{Hdf}\end{eqnarray}
where $v_{\pm}^k=v_{\pm}^o\pm v^e_{\pm}e^{-i\omega k}$. The final diagonal form
(\ref{Hdf}) is obtained by means of a Bogoliubov transformation
${c'}^\dagger_{k\sigma}=\sum_{\nu=\pm}U_{k\sigma}^\nu
a^\dagger_{k\nu}+V_{k\sigma}^\nu a_{-k\nu}$ determined through the
diagonalization of $4\times 4$ blocks
\begin{equation}{\cal H}_k=\left(\begin{array}{cccc}b^o&-v_+^k&0&-v_-^k\\
-\bar{v}_+^k&b^e&\bar{v}_-^k&0\\0&v_-^k&-b^o&v_+^k\\
-\bar{v}_-^k &0&\bar{v}_+^k&-b^e
 \end{array}\right)\end{equation}
whose eigenvalues are $\pm\lambda_k^+$, $\pm\lambda_k^-$, with
\[{\lambda_{k}^\pm}^2=\Delta\pm\sqrt{\Delta^2-
|b^ob^e-(v_+^k+v_-^k)(\bar{v}_+^k-\bar{v}_-^k)|^2} \] and
$\Delta=\frac{(b^o)^2+(b^e)^2}{2}+|v_+^k|^2+|v_-^k|^2$. Care should be taken to
select the correct signs of $\lambda_k^\pm$ in order that the vacuum of the
operators $a_{k\nu}$ has the proper $S_z$-parity and represents the lowest
state for this parity.

The spin correlations in the lowest states for each $z$-parity can then be
obtained from the ensuing basic fermionic contractions $f_{ij}=\langle
c^\dagger_i c_j\rangle-\half \delta_{ij}$, $g_{ij}=\langle c^\dagger_i
c^\dagger_j\rangle$, which can be directly obtained from the inverse Fourier
transform of $\langle {c'}^\dagger_{k\sigma}{c'}_{k\sigma'}\rangle=\sum_\nu
V_{k\sigma}^\nu \bar{V}_{k\sigma'}^\nu$, $\langle
{c'}^\dagger_{k\sigma}{c'}^\dagger_{-k\sigma'}\rangle=\sum_\nu  V_{k\sigma}^\nu
U_{-k\sigma'}^\nu$. We then obtain, through the use of Wick's theorem, $\langle
s^z_i\rangle=f_{ii}$, $\langle s^z_is^z_j\rangle=f_{ii}f_{jj}-f^2_{ij}+g_{ij}^2$, 
and $\langle s_{i}^+s_{j}^\mp\rangle=\frac{1}{2}[{\rm det}(A^+_{ij})\pm
{\rm det}(A^-_{ij})]$, where $A^\pm_{ij}$ are $(i-j)\times(i-j)$ matrices of
elements $2(f+g)_{i+p+ ^0_1,i+q+ ^1_0}$.

\end{document}